\documentclass{article}
\usepackage{graphicx}
\graphicspath{%
    {converted_graphics/}
    {/}
}
\begin{document}

\begin{center}
{\huge Toward a Comprehensive Model of}\vskip3pt

{\huge Snow Crystal Growth Dynamics:}\vskip9pt

{\LARGE 1. Overarching Features and Physical Origins}\vskip16pt

{\Large Kenneth G. Libbrecht}\vskip4pt

{\large Department of Physics, California Institute of Technology}\vskip-1pt

{\large Pasadena, California 91125}\vskip-1pt

\vskip18pt

\hrule\vskip1pt \hrule\vskip14pt
\end{center}

\textbf{Abstract.} We describe a comprehensive model for the formation and
morphological development of atmospheric ice crystals growing from water
vapor, also known as snow crystals. Our model derives in part from empirical
measurements of the intrinsic ice growth rates as a function of temperature
and supersaturation, along with additional observations and analyses of
diffusion-driven growth instabilities. We find that temperature-dependent
conformational changes associated with surface melting strongly affect layer
nucleation dynamics, which in turn determines many snow-crystal
characteristics. A key feature in our model is the substantial role played
by structure-dependent attachment kinetics, producing a growth instability
that is largely responsible for the formation of thin plates and hollow
columnar forms. Putting these elements together, we are able to explain the
overall growth behavior of atmospheric ice crystals over a broad range of
conditions. Although our model is complex and still incomplete, we believe
it provides a useful framework for directing further investigations into the
physics underlying snow crystal growth. Additional targeted experimental
investigations should better characterize the model, or suggest
modifications, and we plan to pursue these investigations in future
publications in this series. Our model also suggests new avenues for the
continued exploration of ice surface structure and dynamics using molecular
dynamics simulations.

\section{Introduction}

\begin{figure}[t] 
  \centering
  \includegraphics[width=4.8in,keepaspectratio]{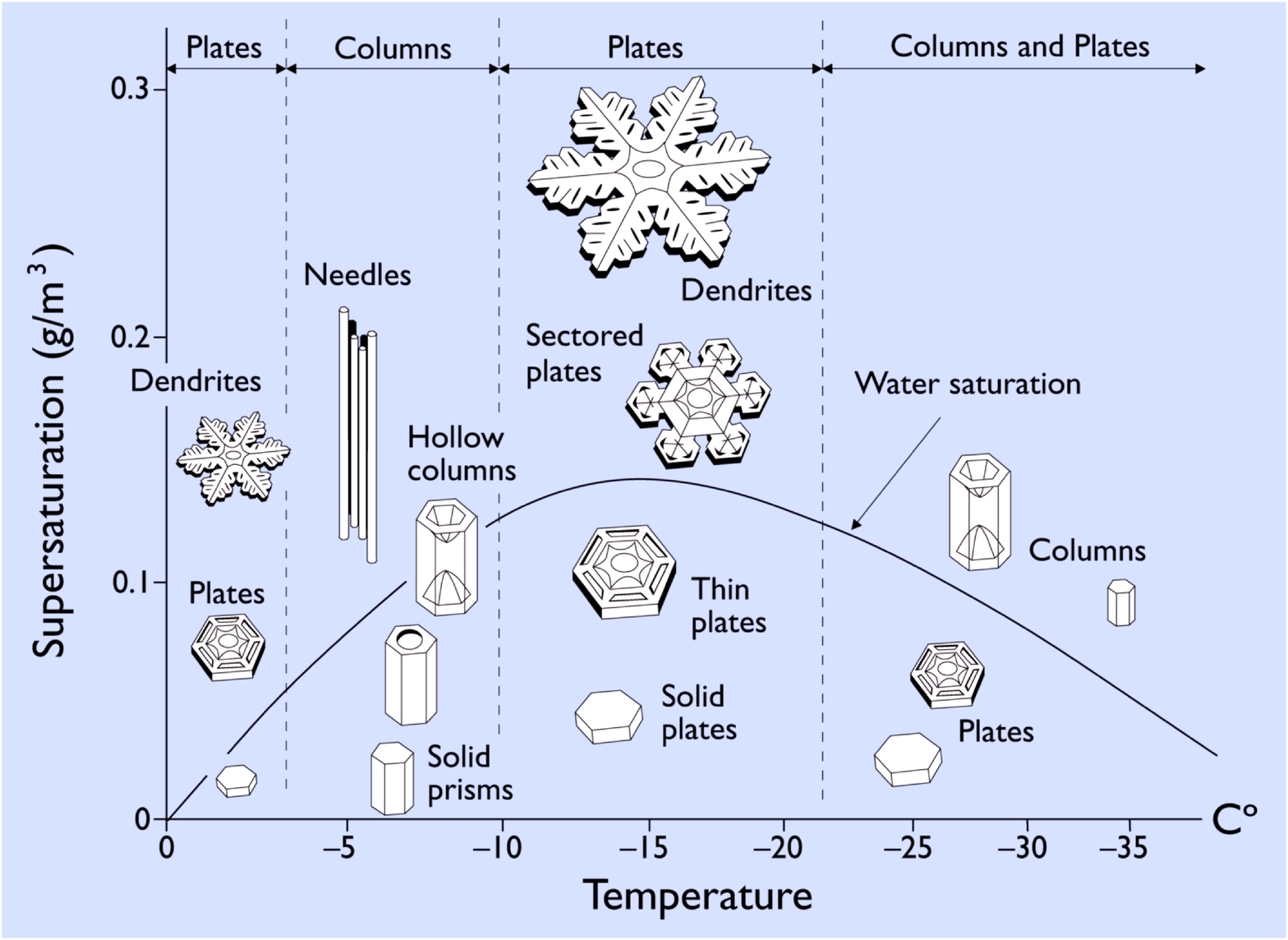}
  \caption{The Nakaya snow crystal
morphology diagram, showing different types of snow crystals that grow in
air at atmospheric pressure, as a function of temperature and water vapour
supersaturation relative to ice. The water saturation line gives the
supersaturation of supercooled water, as might be found within a dense
cloud. Note that the morphology switches from plates ($T\approx -2$ C) to
columns ($T\approx -5$ C) to plates ($T\approx -15$ C) to predominantly
columns ($T<-30$ C) as temperature is decreased. Temperature mainly
determines whether snow crystals will grow into plates or columns, while
higher supersaturations generally produce more complex structures.}
  \label{morph}
\end{figure}

Laboratory observations of snow crystals dating back to the 1930s have
revealed a complex dependence of growth morphologies on temperature and
supersaturation \cite{libbrechtreview, nakaya}. Under common atmospheric
conditions, for example, ice crystals typically grow into thin plate-like
forms near -2 C, slender columns and needles near -5 C, thin-walled hollow
columns near -7 C, very thin plates again near -15 C, and columns again
below -30 C. In addition, morphological complexity generally increases with
increasing supersaturation at all temperatures. These observations are often
summarized in the well-known Nakaya morphology diagram, and one example is
shown in Figure \ref{morph} \cite{libbrechtreview}. Extensions to lower
temperatures, as well as more detailed morphological studies, can be found
in the literature \cite{hallett, fukuta, fukuta1}.

Although the overall features and morphological transitions seen in the
Nakaya diagram have been well established empirically, a basic physical
explanation of why snow crystals exhibit this growth behavior has been
surprisingly elusive. In particular, the fact that snow crystals alternate
between plate-like and columnar forms as a function of temperature has been
an outstanding problem for nearly 75 years. The purpose of this paper is to
present a new model for ice growth from water vapor that includes physical
mechanisms that can explain the observed morphological behavior. Although
our model is not simple, and details are still missing, we believe it
provides a reasonable overarching picture of the various physical processes
that define the growth dynamics of atmospheric snow crystals.

Progress toward explaining how complex structures arise spontaneously during
solidification has generally been hard won. The problem is exacerbated by
the fact that numerous physical processes are often involved over many
length scales, and these work in concert to produce the observed structures.
Identifying and characterizing the most important of these processes,
quantifying them with suitable numerical algorithms, and merging the pieces
into a comprehensive model of structure formation has been an ongoing
process for many decades.

It has been known since the mid-1960s, for example, that dendritic
structures arise from the Mullins-Sekerka instability during
diffusion-limited growth \cite{mullins}. Producing a quantitative model of
this process required substantial theoretical effort, however, culminating
in the development of solvability theory during the 1980s \cite{dendrites,
brener}. With this we learned that the overall branching scale is set by
seemingly minor anisotropies in the surface dynamics. The surface energy
anisotropy plays the key role in the case of solidification from the liquid
phase, while for solidification from gaseous precursors the anisotropy in
the surface attachment kinetics is typically the more dominant factor \cite%
{libbrechtreview, kgldends}.

Numerical models of diffusion-limited growth were developed about the same
time as solvability theory, including front-tracking and phase-field
techniques \cite{fronttracking, phasefield}. These methods have enjoyed
considerable success in reproducing dendritic structures arising during
solidification from the melt, such as in metallurgical systems or ice growth
from liquid water. These systems work well computationally in part because
the material anisotropies are typically quite small, with surface energy
differences of perhaps a few percent between faceted and non-faceted
surfaces. For such systems the growth structures are generally smooth and
free of sharp corners or edges, and numerical models tend to be robust and
stable.

In growth from the vapor phase, anisotropies in the surface attachment
kinetics are often very large, resulting in dendritic structures that are
not smooth with continuous derivatives, but are instead strongly faceted
with sharp edges. In this case numerical instabilities can be problematic,
so considerably more care is required to produce stable growth models \cite%
{garcke}. As a result, the usual numerical techniques used for studying
diffusion-limited growth have not yet been able to produce satisfactory snow
crystal structures from reasonable physical inputs.

In 2008-9, Gravner and Griffeath developed cellular automata (CA) techniques
that avoided the numerical instabilities that affected other methods \cite%
{gg0, gg}. These CA models have generated full three-dimensional dendritic
structures that reproduced many characteristics of natural snow crystals,
including growth forms that are both branched and faceted, with sharp edges.
Cellular automata models incorporating more physically derived rules have
since been demonstrated \cite{cakgl}, and with suitable inputs it now
appears possible, at least in principle, to realize a numerical model that
can accurately reproduce snow crystal growth rates and morphologies at all
temperatures and supersaturations.

Analysis of diffusion-limited growth using these theoretical and
computational tools has yielded numerous insights into the dynamics of
structure formation. For example, solvability theory nicely explains why the
tip velocity of a growing dendritic structure depends linearly on
supersaturation for solidification from vapor, while a quadratic dependence
on undercooling is typical for growth from the liquid phase \cite%
{libbrechtreview, kgldends}. In addition, scaling relationships in
diffusion-limited growth models provide an explanation for the increase in
structural complexity that accompanies decreasing vapor diffusion rates \cite%
{cakgl, gonda}.

For snow crystal growth, these theoretical considerations generally explain
why morphological complexity increases with supersaturation, crystal size,
and background gas pressure. Thus the observed variation along the
supersaturation axis on the morphology diagram in Figure \ref{morph} is
fairly well understood, at least at a qualitative level. Producing accurate
model crystals using sensible input physics over a range of conditions has
not yet been accomplished, and some unusual dendritic snow crystal
structures may be quite challenging to reproduce \cite{fishbones}.
Nevertheless, diffusion-limited growth -- the underlying physical mechanism
responsible for the formation of much snow crystal structure -- is
reasonably well understood, and satisfactory computational algorithms
describing this process are available.

The ability to make high-fidelity numerical models of dendritic growth is
only the first step, however, since models require inputs. For the snow
crystal case, one of the most important physical inputs comes from the
many-body interactions that determine how water molecules are incorporated
into the crystalline lattice, referred to as the \textit{surface attachment
kinetics. }For a rough surface, this incorporation is essentially
instantaneous for all molecules that strike the surface. But the attachment
to faceted surfaces is much slower, and in snow crystal formation the
attachment kinetics are often as important as diffusion in governing the
morphological development. Therefore, understanding the growth of faceted
ice surfaces is essential for any snow crystal model.

Some relevant physical processes that occur during ice crystal growth from
vapor include the adsorption and evaporation of surface admolecules, their
diffusion along the surface, the nucleation of new molecular layers, and the
Gibbs-Thomson effect. Near the melting point one must also consider the high
vapor pressure of ice, along with temperature-dependent surface structural
effects associated with surface melting \cite{dash}. Characterizing the
many-body physics of these various effects remains a significant challenge.

The surface attachment kinetics for ice are sufficiently complex that
crystal growth rates cannot yet be determined from \textit{ab initio}
molecular dynamics modeling, and it appears that this direct approach will
be unfeasible for the foreseeable future. Nevertheless, one need not track
every molecule to create a reasonable physical picture of the
crystallization process, or to realize a useful parameterization of crystal
growth rates. Another route is to combine basic theoretical considerations
with empirical data to create a more simplified crystal growth model,
hopefully one that explains the morphology observations with as much
quantitative accuracy as possible.

Statistical models of crystal growth dynamics have been developed over many
decades also, and some notable early advances in this area include the
formulation of classical nucleation theory, as well as the role of screw
dislocations in crystal growth (for example, see \cite{saito}). The former
is especially important in snow crystal growth, while the effects of
dislocations are often negligible \cite{intrinsic}. Crystal growth theory
has generally enjoyed much success describing many faceted materials,
especially cases where the surface structure is very stable and the
admolecule dynamics are especially simple.

Unfortunately, the ice surface has a high equilibrium vapor pressure, which
means that the surface is in a highly dynamical state. Under typical
conditions, for example, ice molecules evaporate from and redeposit onto the
surface at rates of order one monolayer per microsecond, while crystal
growth rates can easily be 4-6 orders of magnitude slower. Add the
conformational changes associated with surface melting, and it is easy to
see why understanding the molecular physics of ice crystal growth presents
some unique theoretical challenges.

Because the ice surface dynamics are only poorly understood, snow crystal
growth models to date have been quite speculative in nature. Some early
models in the 1960s focused on surface diffusion rates and step growth
velocities, and how these depended on temperature for the principal facet
surfaces (for a review, see \cite{hobbs}). This approach now appears to be
incorrect, as subsequent experiments have shown that the nucleation of new
molecular layers on the principal facet surfaces is the rate-limiting step
in their growth. Recent evidence for this will be discussed below.

One of the best previous attempts to create a comprehensive and physically
motivated model of snow crystal growth was that presented by Kuroda and
collaborators in the 1980s \cite{kuroda, kkuroda, fukuta2}. This model
describes a qualitative picture of the influence of surface melting on
crystal growth rates, and how temperature-dependent surface melting could
explain the observed spectrum of growth morphologies. Subsequent
experimental data have not supported this model, but the underlying
assumption that surface melting plays an important role appears to be sound.

More recent experimental and modeling efforts have convincingly demonstrated
that layer nucleation plays an important role in the formation of faceted
ice surfaces (see \cite{nelson, intrinsic} and references therein), as had
been assumed in the Kuroda model and in some earlier models. Moreover, it
has certainly been demonstrated that bulk diffusion through the surrounding
gas is a key factor influencing growth rates \cite{westbrook}. To date,
however, no model has convincingly explained, even in a qualitative fashion,
the temperature dependence seen in the Nakaya morphology diagram.

Our approach to this problem has been to use quantitative growth
measurements to guide our model building efforts, and to use numerical CA
modeling to compare data with calculated growth rates and morphologies over
a range of conditions. We first observed growth rates at reduced background
pressures to nearly eliminate bulk diffusion effects, thus giving
measurements of the intrinsic growth rates of the ice facet surfaces. With
these data in hand, we then examined growth rates and morphologies at normal
atmospheric pressures, thereby allowing a quantitative analysis of
diffusion-driven growth instabilities. Our result is a comprehensive model
that agrees with the most reliable growth data and explains the general
features in the snow crystal morphology diagram.

Although our model is not without speculative elements, we feel it is a
significant improvement over previous attempts. The model provides a useful
overarching picture of the underlying physics, as well as a reasonable
framework for continued discussion of this problem. For example, we have
found that a key quantity in our model is the step energy $\beta $
associated with the edge of a molecular terrace on the facet surfaces. As $%
\beta $ is a molecular-scale, equilibrium property of the crystal surface,
its calculation should be much easier than full dynamical calculations of
crystal growth rates. Thus it may be possible to use current \textit{ab
initio} modeling techniques to examine some aspects of snow crystal growth
down to the molecular scale.

We have also found that our model is very useful for suggesting targeted
experimental investigations that can both test the model predictions and
better determine the model parameters. In such a capacity, we believe it
likely that continued investigation of this model will be fruitful, and may
yield new insights into ice growth dynamics in different circumstances. We
are currently working on several such experiments, and the results from
these will be presented in future publications in this series.

\section{Intrinsic Growth of the Principal Facets}

\begin{figure}[t] 
  \centering
  \includegraphics[width=3.2in, keepaspectratio]{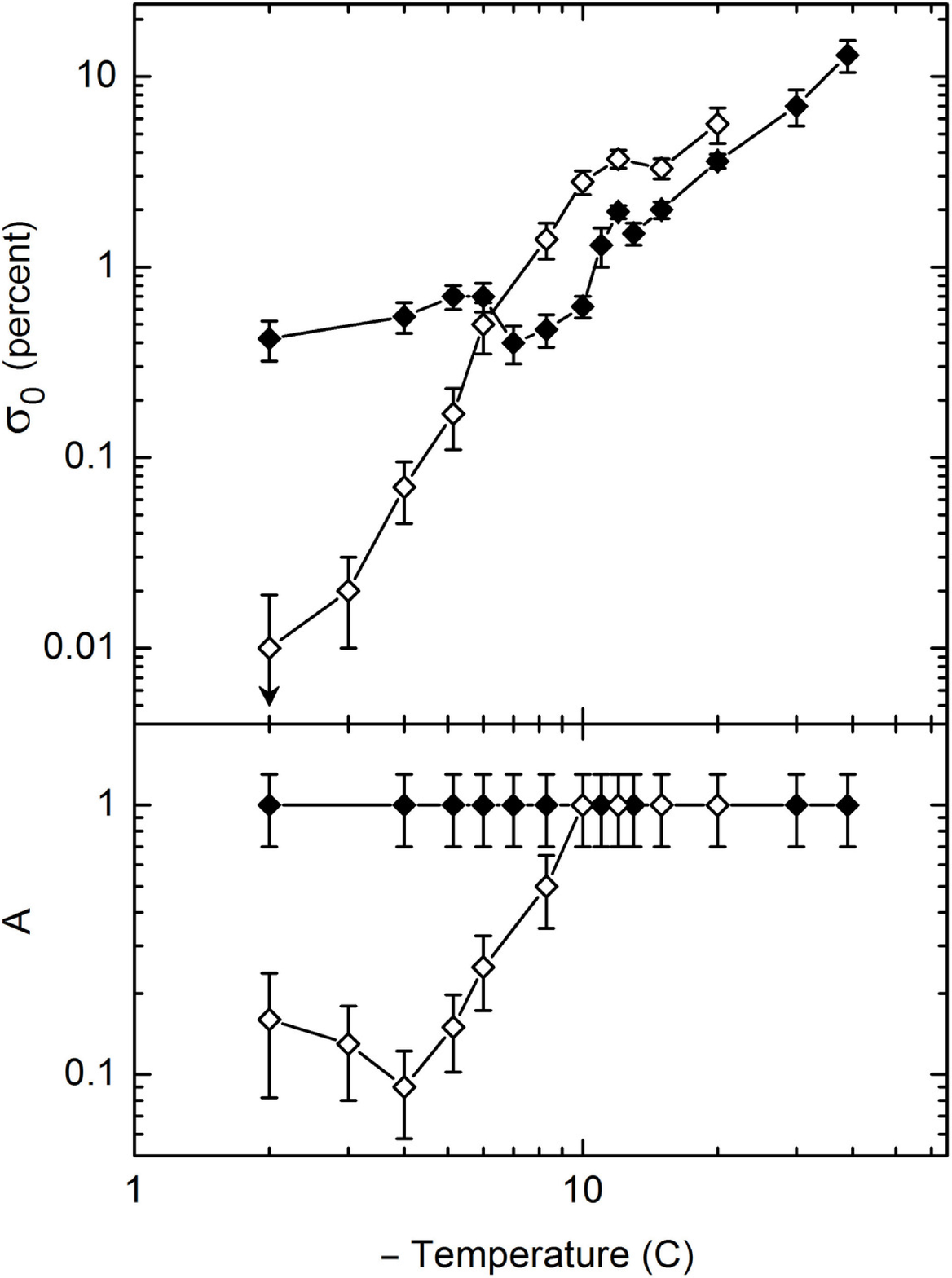}
  \caption{Measurements of the intrinsic
growth rates of the principal ice crystal facets. The growth velocity normal
to the surface is described by $v=\protect\alpha v_{kin}\protect\sigma %
_{surf}$, where $\protect\sigma _{surf}$ is the supersaturation at the
surface and the attachment coefficient is parameterized with $\protect\alpha %
(T,\protect\sigma _{surf})=A\exp (-\protect\sigma _{0}/\protect\sigma %
_{surf}).$ The solid points show the measured $A(T)$ and $\protect\sigma %
_{0}(T)$ for the basal facets, while the open points show measurements of
the prism facets, from \protect\cite{intrinsic}.}
  \label{intrinsic}
\end{figure}

An important step toward understanding snow crystal growth dynamics is to
quantify the growth rates of the principal facet surfaces as a function of
temperature and supersaturation. Following the notation in \cite%
{libbrechtreview}, we parameterize the surface growth velocities using $%
v=\alpha v_{kin}\sigma _{surf},$ where $v$ is the perpendicular growth
velocity, $v_{kin}(T)$ is a temperature-dependent \textquotedblleft
kinetic\textquotedblright\ velocity derived from statistical mechanics, and $%
\sigma _{surf}$ is the water vapor supersaturation relative to ice at the
growing surface. The attachment coefficient $\alpha ,$ which depends on $T$, 
$\sigma _{surf},$ and other factors, encapsulates the attachment kinetics
governing crystal growth at the crystal/vapor interface. From the definition
of $v_{kin},$ we must have $\alpha \leq 1.$

For the simplest case -- the growth of an infinite, clean, dislocation-free
faceted ice surface in near equilibrium with pure water vapor at a fixed
temperature -- the attachment coefficient is well defined and we must have a
unique $\alpha (\sigma _{surf},T)$ for each facet surface. We refer to the $%
\alpha (\sigma _{surf},T)$ for the two principal facets in this ideal case
as the \textquotedblleft intrinsic\textquotedblright\ attachment
coefficients.

We determined $\alpha (\sigma _{surf},T)$ through a lengthy series of
measurements of crystals growing on a substrate at low background pressure.
Experimental details, the resulting data and analysis, and references to
prior work measuring $\alpha (\sigma _{surf},T),$ can be found in \cite%
{intrinsic}. Over the temperature range $-2$ C $>T>-40$ C, our growth data
are well described by a dislocation-free layer-nucleation crystal growth
model, which we parameterize using $\alpha (\sigma _{surf},T)=A\exp (-\sigma
_{0}/\sigma _{surf})$. The measured parameters $A(T)$ and $\sigma _{0}(T)$
for the basal and prism facets are shown in Figure \ref{intrinsic}.

The data in Figure \ref{intrinsic} are the most recent, most extensive, and
we believe the most accurate measurements to date of the intrinsic $\alpha
(\sigma _{surf},T)$ for the principal facets of ice. Several sources of
potential systematic errors were examined with greater care than had been
done in previous experiments, including effects from substrate interactions
and bulk diffusion, and these errors were largely eliminated. Possible
errors from chemical contamination of the ice surface were examined as well 
\cite{chemical}. Within the stated error estimates, therefore, we assume
that the data in Figure \ref{intrinsic} represent empirical fact that must
be incorporated within our model of snow crystal growth dynamics.

\begin{figure}[ht] 
  \centering
  \includegraphics[width=4in,keepaspectratio]{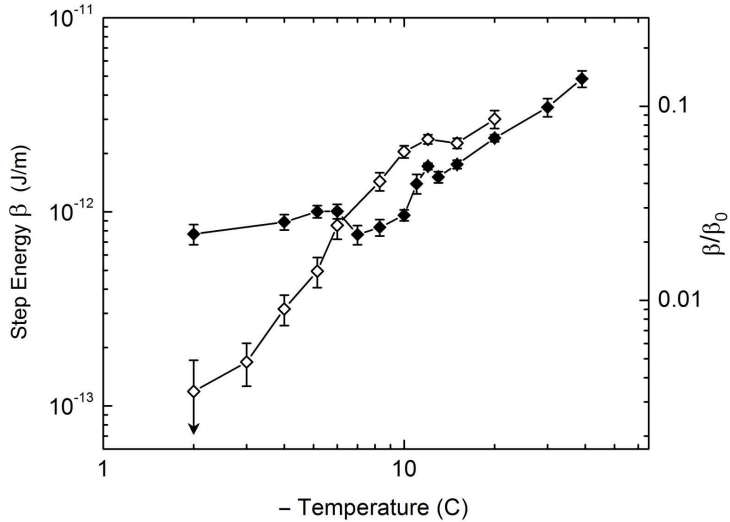}
  \caption{The step energy $\protect\beta %
(T)$ associated with the edge of a molecular terrace for the two principal
facets of ice. This quantity was derived from the measured $\protect\sigma %
_{0}(T)$ in Figure \protect\ref{intrinsic} using classical nucleation theory 
\protect\cite{intrinsic}. For comparison we also show $\protect\beta $
relative to $\protect\beta _{0}=\protect\gamma a,$ where $\protect\gamma $
is the ice surface energy and $a$ is the molecular step height, as described
in the text.}
  \label{beta}
\end{figure}

\subsection{Physical Origins - I}

Our understanding of the detailed molecular structure and dynamics of the
ice surface is not sufficient to provide a full physical explanation of the
measurements displayed in Figure \ref{intrinsic}, and \textit{ab initio}
molecular dynamics simulations cannot accurately calculate ice growth rates
directly \cite{moldymice1, moldymice2, moldym1}. However, the problem is
greatly simplified with the realization that classical nucleation theory
dictates that the parameter $\sigma _{0}(T)$ derives from the step energy $%
\beta (T)$ associated with the edge of a molecular terrace on the facet
surface \cite{saito}. Using this theory, we are able to convert the
measurements to give $\beta (T)$ for both facets, as shown in Figure \ref%
{beta} \cite{intrinsic}.

By taking advantage of nucleation theory, we are thus able to reduce the
problem of crystal growth to one of determining of the step energy. This
represents a major simplification of the problem, since crystal growth is a
complex dynamical process, while the step energy is a basic equilibrium
property of the ice surface. Although $\beta (T)$ has not yet been
calculated from existing ice surface models, this goal may be reached in the
near future. As part of our model, we suggest the following qualitative
picture of how temperature-dependent surface melting can explain much of the
detail seen in Figure \ref{beta}.

Beginning at the lowest temperatures in Figure \ref{beta}, we see that $%
\beta (T)$ increases monotonically with decreasing temperature for both
facets below $T=-15$ C. For extremely low temperatures, we would expect that
the ice surface structure would approach that of a simple molecular layer
model (essentially a basic Kossel crystal \cite{kossel}), so the step energy
would approach $\beta \approx \beta _{0}=\gamma a,$ where $\gamma $ is the
ice surface energy and $a$ is the step height. This follows because in a
simple layer model the increased surface area associated with a step is $aL$%
, where $L$ is the step length, so the increase in effective surface energy
is $\gamma aL,$ giving $\beta \approx \gamma a.$

This reasoning suggests that $\beta \rightarrow \beta _{0}$ at very low
temperatures, and we consider this to be a prediction of our model.
Additional experiment and theory could be used to confirm this hypothesis.
As temperature is increased, we would expect that thermal fluctuations would
influence the structure of the terrace step, causing a surface
reconstruction that effectively smooths out the step, reducing $\beta .$
Thus we would expect $\beta $ to decrease monotonically with increasing
temperature, with $\beta /\beta _{0}<1,$ and this behavior is seen in Figure %
\ref{beta} for $T<-15$ C.

Because the step energy $\beta $ is a molecular-scale equilibrium property
of the ice surface, similar in this regard to the surface energy $\gamma ,$
it appears promising that $\beta $ may be calculable using molecular
dynamics simulations. The low-temperature region is probably the most
amenable to such investigations, since for $T<-15$ C it appears that the
surface structure is relatively simple and not complicated by substantial
surface melting. If the measured approximate power-law behavior in $\beta (T)
$ could be reproduced for the low-temperature region, this would give one
confidence that our characterization of the step energy is a reasonable one,
as well as providing physical insights into the underlying molecular origins
of $\beta (T).$

Continuing up in temperature in Figure \ref{beta}, we see unusual features
in both $\beta _{basal}$ and $\beta _{prism}$ at $T\approx -12$ C. A sharp
peak in $\beta _{basal}(T)$ is seen at this temperature, as is a broader
peak (little more than a shoulder) in $\beta _{prism}(T)$. Both peaks are
robust features in our data, and we suggest that both are associated with
the onset of significant surface melting in the ice surface. By
\textquotedblleft onset\textquotedblright\ in this context, we mean the
temperature at which surface melting first begins to significantly affect
the crystal growth dynamics. This interpretation is consistent with some
other measurements of surface melting in ice \cite{dosch}, although in
general our theoretical understanding of surface melting in ice and other
materials is rather poor \cite{dash}.

Exactly why the onset of surface melting produces peaks in $\beta (T)$ near $%
-12$ C is not known. We can only speculate that the observed behavior may
result from complex many-body dynamics at this temperature, when the top
molecular layers are just beginning to become disordered. If this is true,
then explaining these $-12$ C features in detail may be quite challenging.
Their effect on snow crystal growth dynamics, however, appears to be fairly
small compared to the overall behavior in $\beta (T)$ for the two facets.

Focusing now on the prism facet, we see in Figure \ref{beta} that $\beta
_{prism}(T)$ declines precipitously with increasing temperatures above $%
T\approx -10$ C. At these temperatures, surface melting produces a
substantial quasi-liquid layer (QLL) in ice, so the ice/vapor interface
becomes replaced by an ice/QLL/vapor interface. The data then suggest that
surface reconstruction further smooths the step edge on the prism facet,
thereby reducing $\beta _{prism}$ with increasing temperatures. Here again
the situation seems ripe for additional investigation with molecular
dynamics simulations. Investigations using \textit{ab initio} calculations
have yielded considerable insights into ice surface melting \cite%
{moldymice1, moldymice2, moldym1}, and and it seems promising that examining
step energies with these methods may well explain some of the features seen
in Figure \ref{beta}.

For the basal facet, we see that $\beta _{basal}(T)$ initially drops for
temperatures above $T\approx -10$ C, but then the trend reverses and $\beta
_{basal}(T)$ exhibits a brief upward jump with increasing temperature for $%
T\approx -6.5$ C, finally leveling off for $T>-4$ C. To explain this
behavior, we borrow a key feature from the Kuroda model and suggest that
extensive surface melting at higher temperatures yields a sharp ice/QLL
interface on the basal facet. We suggest that this interface is accompanied
by a $\beta _{basal}$ that approaches some constant value as the temperature
approaches the melting point and the QLL thickness diverges. In this
behavior we see an important difference between the basal and prism facets
of ice -- near the melting point, the nucleation barrier essentially
vanishes on the prism facet (its effects become negligible in the crystal
growth dynamics) while remaining substantial on the basal facet.

\medskip 

\subsubsection{Connection to Ice Growth from Liquid Water}

This high-temperature picture is consistent with observations of ice growing
from liquid water, where one typically sees faceting on the basal facet but
no faceting on the prism facet \cite{maruyama}. This makes sense given our
understanding that the QLL thickness diverges near the melting point, so one
expects similar nucleation barriers at the ice/water interface and the
ice/QLL interface when the QLL thickness is large.

Another prediction from our model, therefore, is that there should be a
correspondence between $\beta _{basal}$ near the melting point as determined
from two different measurement strategies -- using ice growth from water
vapor, and using ice growth from liquid water. To our knowledge, this
correspondence has not been investigated to date. Experiments on ice growth
from liquid water have not yet determined $\beta _{basal}$ accurately \cite%
{maruyama}, mainly because of interference from heat diffusion effects. But
it appears that an accurate measurement of $\beta _{basal}$ for the liquid
case could be achieved with a targeted experimental investigation. Examining
the correspondence between growth from water vapor near 0 C and growth from
liquid water at low supercoolings may well yield additional insights into
the attachment kinetics in both cases.

\medskip 

\subsubsection{Surface Diffusion and $A(T)$}

Early attempts to model the temperature dependence in snow crystal growth
focused on the 2D diffusion of water admolecules on the ice surface, as
mentioned above and discussed at length in \cite{hobbs}. The measurements in
Figure \ref{intrinsic} tell us, however, that surface diffusion plays a
relatively minor role compared to nucleation dynamics. From crystal growth
theory in a multinucleation model (see \cite{saito}), we find that the
growth parameter $\sigma _{0}(T)$ is determined mainly by the step energy $%
\beta (T),$ while the prefactor $A(T)$ is determined mainly from the surface
diffusion of admolecules and their attachment to the terrace edges. (Note
that we have ignored a factor of $\sigma _{surf}^{1/6}$, as this factor is
typically inconsequential compared to the nucleation factor $\exp (-\sigma
_{0}/\sigma _{surf})$.) Since $A(T)$ derives from an extrapolation of the
growth data to high $\sigma _{surf}$ \cite{intrinsic}, and is further
subject to the constraint that $\alpha \leq 1,$ we see that fast surface
diffusion yields the limit $A(T)=1,$ as is observed on the prism facet for $%
T\leq -10$ C and on the basal facet at all temperatures measured.

From Figure \ref{intrinsic} we see that $A_{prism}(T)$ is less than unity
only for $T>-10$ C, when surface melting is well developed. We suggest that $%
A_{prism}\ $is reduced in this region because of a growth impedance at the
ice/QLL interface, although the microphysics underlying this is not known.
Additional investigation of this behavior, and comparison with other crystal
systems, would likely be fruitful.

At temperatures near $T=-2$ C, the nucleation barrier on the prism facet is
quite low, so the main impediment to the intrinsic growth of the prism facet
comes from the fact that $A_{prism}<1.$ Because of this, the growth of the
prism facet is slower than a rough surface, which results in faceting. In
growth from liquid water we see no faceting of the prism facet, however.
This suggests another prediction of our model, that $A_{prism}(T\rightarrow
0)\approx 1,$ owing to the correspondence between growth from liquid water
and growth when the QLL is thick. Extending the data in Figure \ref%
{intrinsic} to higher temperatures, although challenging experimentally,
could test this prediction.

\subsection{Further Investigations of Intrinsic Growth Rates}

Although the data in Figure \ref{intrinsic} represent a significant step
forward in understanding the intrinsic growth behavior of the principal
facets of ice, extending these measurements to both lower and higher
temperatures would be beneficial. The low-temperature regime is probably the
easiest to model with molecular dynamics simulations, so better measurements
here could facilitate progress in characterizing the ice surface in the
absence of surface melting. Reproducing the observed $\sigma _{0}(T)$ with 
\textit{ab initio} calculations would be a substantial achievement, even if
only in the low-temperature regime, given our current poor understanding of
the molecular structure and dynamics of the ice surface.

At the high-temperature end of the spectrum, comparing ice growth from water
vapor and from liquid water would also likely be fruitful. Better data are
needed for both systems, and the theory describing the correspondence needs
to be developed. Comparing the ice/water and ice/QLL interfaces would
certainly help develop our knowledge of these interfaces, as well as
improving our overall understanding of surface melting in the thick-QLL
limit.

Although the physical picture we have painted above is imperfect and
somewhat speculative, we believe it provides a useful framework for further
investigations. Our intention in defining this model is in part to create a
plausible qualitative description of the underlying physics that is
consistent with the experimental data. In spite of its shortcomings, we
believe that our model of the intrinsic growth behavior does make some
useful predictions, described above, as well as suggesting additional
targeted investigations, both experimental and theoretical.

\section{Structure Dependent Attachment Kinetics}

The measured intrinsic growth rates, as parameterized above and displayed in
Figure \ref{intrinsic}, immediately appear to contradict many details in the
Nakaya morphology diagram. For example, we see that $\sigma
_{0,basal}<\sigma _{0,prism}$ at -15 C, implying that $\alpha
_{prism}<\alpha _{basal}$ for all supersaturations at this temperature. This
inequality suggests that columnar prisms would be the preferred growth
morphology, while it is well established that thin plates form at this
temperature.

This discrepancy is explained in our model with the phenomenon of \textit{%
structure-dependent attachment kinetics} (SDAK) \cite{sdak1, sdak2, sdak3},
which causes the attachment coefficient $\alpha (\sigma _{surf})$ to depend
on the mesoscale morphological structure of the ice surface itself. For
example, our SDAK model at -15 C assumes that $\alpha _{prism}$ on a thin
plate edge is higher than the corresponding \textquotedblleft
intrinsic\textquotedblright\ $\alpha _{prism}$ for a large faceted surface.
The increased $\alpha _{prism}$ then reverses the above inequality, yielding 
$\alpha _{prism}>\alpha _{basal}$ and resulting in the growth of thin plates.

More specifically, in our SDAK model the increase in $\alpha _{prism}$ on
the edge of a thin plate arises because $\sigma _{0,prism}$ decreases as the
width of the final molecular terrace on the prism surface approaches
molecular dimensions. Observations indicate that the radius of curvature of
the edge of a thin plate at -15 C is typically $R\approx 0.5$ $\mu $m, so
the width of the last molecular terrace on the prism surface is roughly $%
w\approx (aR)^{1/2}\approx 40a,$ where $a\approx 0.3$ nm is the size of a
water molecule. The fundamental premise of our SDAK model is that surface
melting and perhaps other structural effects will become altered on such a
narrow terrace, affecting the nucleation dynamics. Since the nucleation of
new molecular layers on this thin terrace largely determines the growth rate
of the prism edge, it follows that SDAK effects play an important role in
governing ice growth morphologies.

Here again, our detailed understanding of the ice surface is not sufficient
to define a incontrovertible model of structure dependent attachment
kinetics from first principles. However, growth measurements do demonstrate
the existence of this phenomenon \cite{sdak3}, and these same measurements
can characterize the SDAK effects to some degree. We proceed with the
understanding that our SDAK model is quite preliminary at present.
Additional measurements and theoretical work will be necessary to understand
the importance of SDAK effects on both facets as a function of temperature
to a satisfactory level.

\begin{figure}[t] 
  \centering
  \includegraphics[width=6in,keepaspectratio]{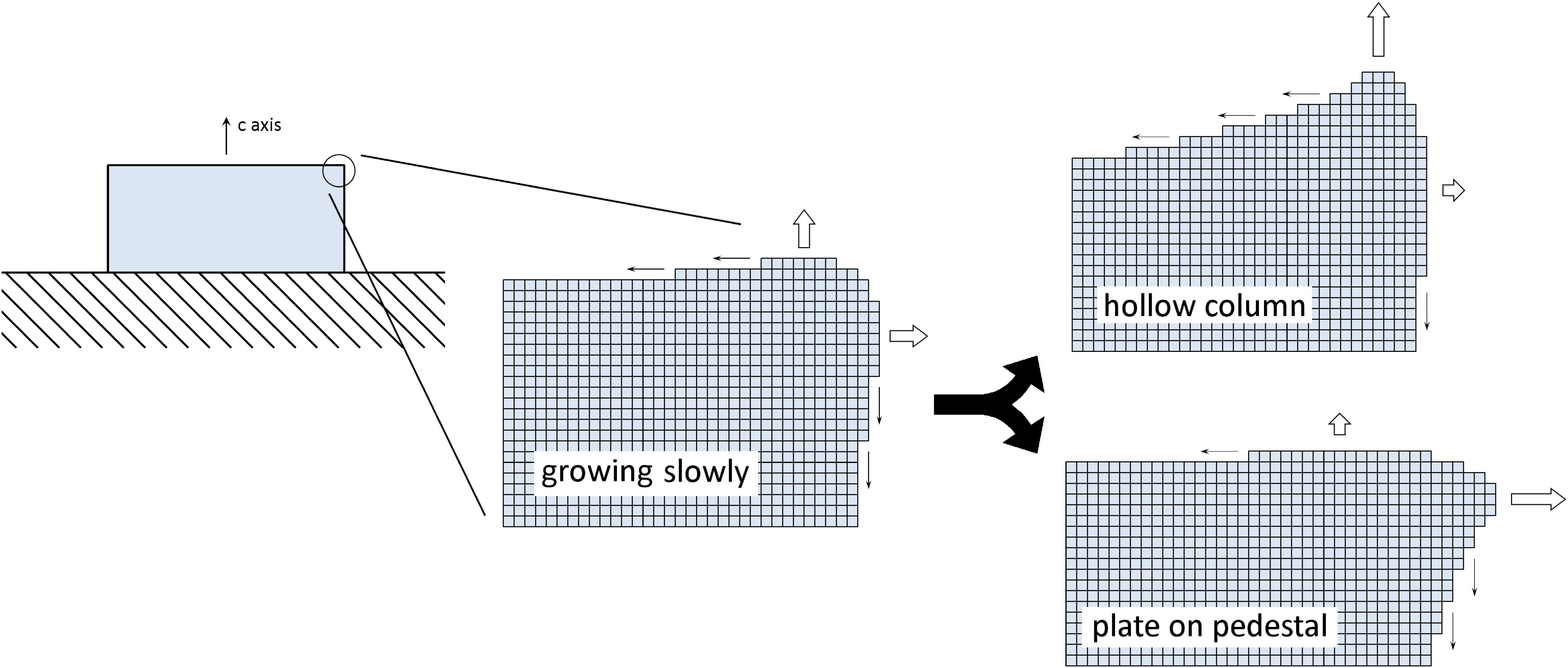}
  \caption{A schematic depiction of the
SDAK instability described in the text. The growth of the corner of a
faceted ice crystal prism (left) is dominated by the nucleation of terraces
on the basal and prism facets (center). If the nucleation rate increases as
the width of the top basal terrace decreases (top right), in keeping with
the SDAK model, then the accelerated growth narrows the basal surface,
accelerating the growth still more. The resulting positive feedback
generates the growth of a hollow columnar crystal. If the same SDAK effect
is more prevalent on the prism facet (lower right), then the instability
leads to the growth of a thin plate from the top edge of the prism.}
  \label{sdak}
\end{figure}

\subsection{An Edge-Enhancing SDAK\ Instability}

A key feature of our SDAK hypothesis is that it leads to an edge-enhancing
growth instability. The essential mechanism is that as a thin edge begins to
form, $\sigma _{0}$ decreases and thus further increases the edge growth
rate. The enhanced growth causes the edge to sharpen, which again increases
the growth rate. This positive feedback yields a growth instability that
enhances the formation of sharp edges.

To examine the SDAK\ instability in more detail, consider the growth of an
initially isometric prism on a substrate, depicted in Figure \ref{sdak}. We
assume the presence of an inert background gas surrounding the crystal, so
the growth is partially diffusion limited. If the crystal is growing slowly
(center diagram in the figure), then molecular terraces nucleate slowly near
each corner of the crystal, where $\sigma _{surf}$ is locally highest, and
steps propagate away from the corner. The corner itself is rounded from the
Gibbs-Thomson effect. For slow growth, this is essentially the standard
model of diffusion-limited faceted crystal growth, resulting in slightly
concave faceted surfaces.

Consider now the highest molecular terrace on either facet next to the
growing corner. As the supersaturation is increased, new terraces nucleate
more frequently, so the average width of the top terrace decreases. As the
SDAK effect reduces $\sigma _{0}$ on narrower terraces, the nucleation rate
increases and in turn the more rapid growth further decreases the width of
the top terrace.

At this point a competition occurs between growth on the basal and prism
facets, as shown in the pair of diagrams on the right side of Figure \ref%
{sdak}. If the SDAK effect preferentially reduces $\sigma _{0}$ on the basal
facet (top right diagram), then the basal growth is especially enhanced.
Because the growth is also diffusion-limited, the fast growth on the basal
facet depletes the water vapor supply from the nearby prism facet. This
decreases the nucleation rate on the prism facet, which causes the average
width of the top terrace to increase, which in turn increases $\sigma
_{0,prism}.$ The combined effect is that $\sigma _{0,prism}$ increases to
essentially its intrinsic value while $\sigma _{0,basal}$ grows ever smaller
as the basal edge grows sharper. The final result is a hollow column
morphology with thin basal edges. Alternatively, the same instability could
favor the prism facets, as seen in the lower right diagram in Figure \ref%
{sdak}. In this case a thin plate-like crystal would form on the isometric
prism, producing a \textquotedblleft plate-on-pedestal\textquotedblright\
morphology \cite{sdak3}.

Note that the SDAK instability nicely explains the abrupt transitions
between plate-like and columnar growth seen in the morphology diagram.
Relatively small changes in the surface attachment kinetics with temperature
can be amplified via the SDAK instability to yield very substantial changes
in the final crystal morphologies.

Note also that the SDAK instability is essentially an extension of the
well-known Mullins-Sekerka instability in diffusion-limited growth \cite%
{mullins}. The latter is well known for producing dendritic branching during
solidification, but alone it does not explain the formation of thin
plate-like or hollow columnar crystals. Thin-edge morphologies require
strong anisotropies in the attachment kinetics -- namely $\alpha _{prism}\gg
\alpha _{basal}$ for thin plates or $\alpha _{prism}\ll \alpha _{basal}$ for
hollow columns. The SDAK instability provides a natural mechanism to
generate these strong anisotropies.

\subsection{Physical Origins - II}

The SDAK effect is difficult to fully evaluate using growth measurements
alone, in part because of the SDAK instability. For example, the strong SDAK
effect on the prism surfaces at -15 C promotes the growth of thin plates at
this temperature, which in turn means that the basal facets are quite large.
If an SDAK effect were present on the basal facets at -15 C, we could not
see it because it is essentially overshadowed by the stronger SDAK effect on
the prism surfaces. The bifurcation into plate-like or columnar growth shown
in Figure \ref{sdak} means that only the dominant SDAK effect can be
characterized by growth measurements at a given temperature.

\begin{figure}[ht] 
  \centering
  \includegraphics[width=3in,keepaspectratio]{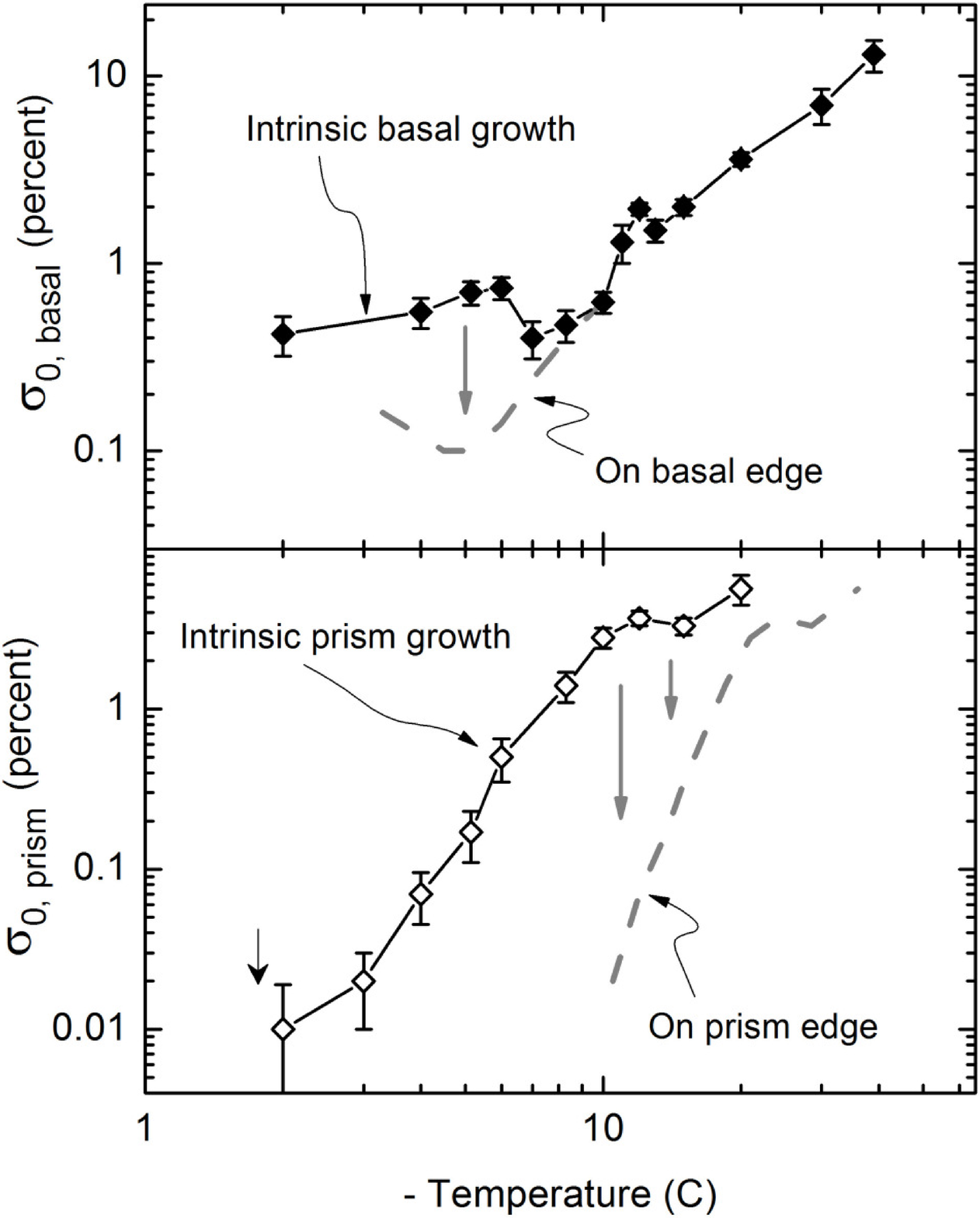}
  \caption{In this graph the solid lines
show the intrinsic $\protect\sigma _{0,basal}(T)$ and $\protect\sigma %
_{0,prism}(T)$ on large facet surfaces, using the data shown in Figure 
\protect\ref{intrinsic}. The dotted lines show our proposed $\protect\sigma %
_{0,basal,SDAK}(T)$ and $\protect\sigma _{0,prism,SDAK}(T)$ when the terrace
width is very small, as a result of the SDAK effect, as described in the
text.}
  \label{combo2}
\end{figure}

We propose that the strong SDAK effect on the prism facet at -15 C is
related to the precipitous drop in $\sigma _{0,prism}$ for temperatures
above -10 C. When a prism facet becomes very narrow at -15 C, we suggest
that the molecular binding in the top terrace is effectively reduced
relative to the binding on a larger facet surface. This reduced binding
causes an increase in surface melting, which in turn causes a decrease in $%
\sigma _{0,prism},$ as one normally sees at higher temperatures. Thus the
function $\sigma _{0,prism}(T)$ is essentially shifted in temperature when
the prism terrace becomes narrow. Although the details of this behavior are
uncertain, we suggest that $\sigma _{0,prism}(T)$ on the top terrace of a
thin edge looks something like the dotted line shown in the lower panel in
Figure \ref{combo2}.

With this picture, we see that the SDAK effect will be less prevalent on a
prism facet below -20 C, because a larger effective shift in temperature
would be needed to substantially reduce $\sigma _{0,prism},$ and such large
shifts are not provided by the SDAK mechanism. Thus in Figure \ref{combo2}
we see that $\sigma _{0,prism,SDAK}\approx \sigma _{0,prism}$ at -20 C. Put
another way, the SDAK effect on the prism facet is most prevalent when the
temperature is close to the drop-off in $\sigma _{0,prism}$ associated with
surface melting. At temperatures below -20 C, reducing the width of the
prism facet is not sufficient to induce surface melting and its accompanying
drop in $\sigma _{0,prism}.$ The end result is that the tendency to produce
thin plate-like crystals is reduced below -20 C, as is observed.

On the other hand, we would expect that the SDAK effect at -10 C on the
prism facet would be substantial, since the intrinsic $\sigma _{0,prism}(T)$
continues to drop at still higher temperatures. However, once $\sigma
_{0,prism}$ is reduced to values of order 0.1 percent, the nucleation
barrier becomes so low that any additional reduction would have little
effect on the growth rates. At the same time, $\sigma _{0,basal}$ drops with
increasing temperature near -10 C as well. When both $\sigma _{0,basal}$ and 
$\sigma _{0,prism}$ are low, then the anisotropy in $\alpha _{prism}/\alpha
_{basal}$ will decrease, so thin plates no longer form. Putting the pieces
together, this qualitatively explains why the formation of thin plates peaks
at temperatures around -15 C, as seen in the morphology diagram.

Note that this overall physical picture suggests that there might exist a
weaker SDAK effect on the basal facet at -15 C, since $\sigma _{0,basal}$
also drops with increasing temperature. However, as mentioned above, because
the SDAK instability favors the growth of thin plates at this temperature,
the SDAK effect on the basal facet has little influence on the overall
crystal growth morphology. Because this weaker SDAK effect on the basal
surface at -15 C is quite uncertain, and moreover it is overshadowed by the
SDAK effect on the prism surface at this temperature, is was not included in
Figure \ref{combo2}. Nevertheless it may exist, and it may be measurable by
other means, so we suggest it as a possibility for future investigation.

At temperatures near -5 C, our model includes an SDAK instability favoring
rapid basal growth and the formation of needles and hollow columns. However,
the underlying physical mechanism is quite different from the SDAK effect on
the prism surface near -15 C. In our discussion of $\sigma _{0,basal}(T)$
above, we suggested that the formation of a relatively sharp ice/QLL
interface on the basal facet caused a positive $d\sigma _{0,basal}/dT$ at
-6.5 C, and likewise caused $\beta _{basal}$ to be considerably higher than $%
\beta _{prism}$ at higher temperatures. We further propose that the sharp
ice/QLL interface is relatively fragile when the QLL is thin, and therefore
the sharp interface becomes unstable when the width of the basal facet is
reduced to near molecular dimensions.

In the absence of a sharp ice/QLL interface, we would then expect that $%
\sigma _{0,basal}$ would diminish with increasing temperature, so that $%
\sigma _{0,basal}$ is no longer much larger than $\sigma _{0,prism}$. In
other words, we would expect that $\sigma _{0,basal}(T)$ on a narrow basal
terrace would extrapolate from the behavior at -7 C and lower temperatures
as shown in the top panel Figure \ref{combo2}. The details of this
extrapolation are uncertain, but are not terribly important for the present
discussion; all that matters is that the reduction in $\sigma _{0,basal}$ on
a narrow basal terrace near -5 C is sufficient to bring about the SDAK
instability that promotes columnar growth.

\subsection{Further Investigations of Structure Dependent Attachment Kinetics%
}

Our understanding of SDAK effects and the SDAK instability are rather poor
at present, so much work is needed to develop a clearer picture of this
phenomenon. The measurements in \cite{sdak3} present a rather strong case
that the SDAK instability does exist, at least near -15 C, but little has
been done at other temperatures. Our model clearly predicts that the SDAK
instability should be present at temperatures near -5 C, and demonstrating
this experimentally seems quite straightforward. Investigations of growth
behaviors at other temperatures should also be fruitful, and in principle
one could map out the SDAK effects over a range of conditions.

Observations of growth behaviors as a function of supersaturation,
temperature, air pressure, and initial conditions could all provide useful
insights. As was demonstrated in \cite{sdak3}, however, quantitative
comparisons between growth data and numerical models are essential for
making progress in this area. Careful numerical modeling of
diffusion-limited growth, accurate growth measurements, and well-defined
experimental conditions, are all necessary to disentangle the various
processes affecting snow crystal growth rates and morphologies.

\section{Discussion}

The goal of the present paper is not to have the final word on this subject,
but rather to take a significant step forward in defining a comprehensive
physical model of snow crystal growth dynamics. Toward this end we believe
we have identified many key features in such a model, and we have proposed
plausible physical origins for various aspects of the model behavior. Our
model is admittedly complex, and we have used some \textit{ad hoc}
assumptions to fill in aspects that are not otherwise well constrained. On
the plus side, however, our model makes many quantitative predictions that
can be tested with targeted experimental investigations. Moreover, our model
also suggests theoretical investigations, particularly using molecular
dynamics simulations, that may yield important insights. The usefulness of
any model is its predictive power, and time will tell how well the model
stands up to additional scrutiny.

The first important feature in our model is simply that the transport of
water molecules via diffusion through the surrounding background gas plays a
substantial role in guiding the growth and morphological evolution of a snow
crystal. This conclusion is certainly not new, but the recent development of
physically realistic cellular automata methods appears to be a significant
breakthrough in modeling the diffusion-limited growth of crystals that are
both faceted and branched, as are snow crystals.

A careful look at the diffusion equation shows that the growth of snow
crystals under ordinary atmospheric conditions is almost always strongly
affected by diffusion effects. Only for micron-scale crystals growing at
quite low pressures, or for very slowly growing crystals, is this statement
not correct. Moreover, about the only way to model the diffusion-limited
growth of snow crystals with adequate accuracy is by employing numerical
methods. Without careful diffusion modeling, it is extremely difficult, if
not impossible, to experimentally measure snow crystal growth parameters in
a reliable fashion. This basic fact has not been adequately appreciated in
many previous snow crystal growth experiments.

A second key feature in our model is the central role of the surface
attachment kinetics, especially the intrinsic attachment coefficients of the
principal facets, $\alpha _{basal}$ and $\alpha _{prism}.$ A successful
model must first describe how these parameters depend on temperature $T$ and
supersaturation $\sigma _{surface},$ and must provide a reasonable picture
of the underlying physical processes responsible for the observed behaviors
in $\alpha _{basal}(T,\sigma _{surface})$ and $\alpha _{prism}(T,\sigma
_{surface}).$ This task, by no means simple, is the first step toward
defining a comprehensive model of snow crystal growth dynamics.

In our model we rely on the empirical growth data to define the intrinsic
attachment coefficients. These data are well described by a dislocation-free
layer-nucleation crystal growth model, which we parameterize as $\alpha
(\sigma _{surf})=A\exp (-\sigma _{0}/\sigma _{surf})$, and the measured
parameters $A(T)$ and $\sigma _{0}(T)$ for the basal and prism facets are
shown in Figure \ref{intrinsic}. We believe these data are substantially
more reliable than previous attempts, in part because careful attention was
given to understanding systematic experimental effects arising from particle
diffusion, substrate interactions, and other sources.

We attempted to paint a reasonable physical picture explaining the intrinsic
growth behavior in our discussion above. While we believe this picture is
satisfactory, there may well be other interpretations of the data. This is
clearly an area for additional experimentation and theoretical
investigations. The ice surface has a complex structure and it is far from
being in a static state. Understanding in detail how the structure and
dynamics of the ice surface affect crystal growth rates remains a
significant challenge.

A third key feature in our model is the importance of structure-dependent
attachment kinetics, and this is perhaps the most novel and controversial
aspect of what we have presented above. The evidence supporting the SDAK
phenomenon has become quite strong, and we believe it is now difficult to
deny that SDAK effects are present in snow crystal growth, and that they
play a central role in the formation of thin plate-like crystals near -15 C.
The details in what we presented above may be inaccurate at some level, and
the physical origins we put forth for the SDAK effect may require revision,
but we believe that the SDAK effect must be an essential part of any
comprehensive snow crystal growth model. Further experimental and
theoretical investigations into the SDAK effect and the SDAK instability are
likely to provide important new insights into this phenomenon.

Experimental and theoretical efforts focusing on structure formation in ice
over the past several decades have continually pushed the remaining
frontiers ever closer to the molecular scale. The SDAK instability, the
parameterization of the attachment coefficients for the principal facets, as
well as the properties of surface melting, all arise from the detailed
molecular dynamics at the crystal surface. Since much progress has been made
recently in molecular dynamics simulations of the ice surface, it appears
promising that additional investigations along these lines may reveal new
insights into ice growth behavior, and especially why the ice surface
properties vary with temperature as they do. How these advances apply to
other crystal systems, and to our understanding of surface molecular
dynamics in general, remains to be seen.

\end{document}